\def\l{\lambda }  \def\r{\varrho }  \def\s{$\,$}  \def\t{\theta }
\def\o{\omega }
\font\tenyyy=cmcsc10 \def\yyy{\tenyyy}
\documentstyle[12pt]{article}
\newcommand{\simgeq}{\; \raisebox{-0.4ex}{\tiny$\stackrel
{{\textstyle>}}{\sim}$}\;}

\newcommand{\beq}{\begin{equation}}
\newcommand{\beqar}{\begin{eqnarray}}
\newcommand{\eeq}[1]{\label{#1} \end{equation}}
\newcommand{\eeqar}[1]{\label{#1} \end{eqnarray}}
\begin{document}
\centerline{{\huge Classical Analysis of Phenomenological Potentials}}
\centerline{{\huge for Metallic Clusters}}
\medskip
\centerline{{\yyy
W.D.\s Heiss$^{\star}$  and R.G.\s Nazmitdinov$^{\star \star}$
\footnote{on leave of absence from
Joint Institute for Nuclear Research,
Bogoliubov Laboratory of Theoretical Physics, 141980 Dubna, Russia}
 }}
\medskip
\centerline{{\sl
$^{\star}$ Centre for Nonlinear Studies and Department of Physics}}
\centerline{{\sl
University of the Witwatersrand, PO Wits 2050, Johannesburg, South Africa }}
\centerline{{\sl
$^{\star \star}$ Departamento de Fisica Teorica C-XI }}
\centerline{{\sl Universidad Autonoma de Madrid, E-28049, Madrid, Spain}}
\baselineskip 20pt minus.1pt
\begin{abstract}
The classical trajectories of single particle motion in a Woods--Saxon and
a modified Nilsson potential are studied for axial quadrupole deformation.
Both cases give rise to chaotic behaviour when the deformation in the
Woods--Saxon and the $\vec l\,^2$-term in the modified Nilsson potential are
turned on. Important similarities, in particular with regard to the
shortest periodic orbits, have been found.
\end{abstract}
\newpage
\textheight=23cm
\voffset=-2.2cm

Recent experimental results on metallic clusters reporting abundance
variations in mass spectra, ionisation potentials, static polarisabilities
and collective giant dipole resonances, barrier shapes and fragmentation
provide us with striking manifestations of shell structure effects related to
a quantised motion of the valence electrons \cite{He93}. The correspondence
of the electronic shell structure in spherical
clusters to the closing of major quantal
shells \cite{Bj90,Br93,Ma91} caused considerable interest in using nuclear
shell model type calculations for the description of metallic clusters
\cite{Cl85,Ni90,Fr93,Ya93,Bu93}. It turns out that phenomenological potentials
 used traditionally in nuclear physics serve a purpose similar to those
obtained within the Kohn--Sham density--functional method \cite{KS65} if the
relevant parameters are adjusted appropriately. Typical potentials are the
Woods--Saxon (and its various modifications) and the modified Nilsson
potential without spin--orbit term. The considerable lowering of the
computational time due to their simple analytical form renders an analysis
of the stability of large metallic clusters feasible.
Naturally, the shell numbers have to
be larger than the ones used in nuclear physics in accordance with the larger
number of valence electrons considered for metallic clusters. For mesoscopic
objects like clusters, deformations, i.\s e.\s deviations from spherical
symmetry of the potential, is as important as in the nuclear
physics context. Metallic clusters can be seen as a `gift of nature' towards
a deeper understanding of the formation of shell and supershell structure
which is a general feature for any self--consistent theory of independent
particles moving in an average potential.

The existence of shell structures is one of the crucial questions addressed
by previous \cite{Cl85,Ni90,Cl91,Fr93,Bu93} and more recent authors
\cite{Ma93,Le93,Pa93}. Obviously, shell structure in the quantum mechanical
spectrum is associated with periodic orbits in the corresponding classical
problem \cite{BM75,S76}. Furthermore, if the corresponding classical problem
is nonintegrable and displays chaotic behaviour, the shell structure of
the corresponding quantum spectrum is affected depending on the degree of
chaos \cite{G90,H94}. Since the deformed Woods--Saxon potential as well as
the modified Nilsson potential are nonintegrable systems, a classical analysis
of the single particle motion seems to be indicated to shed light on the
corresponding quantum mechanical problem. Results of such analysis are
presented in this paper. This is relevant for two reasons. Since the
Woods--Saxon potential and the Nilsson model are both used with
success in quantum mechanical models for clusters, it is of
interest to look at their similarities in the classical context; in fact
their similarities are not obvious at first glance. Furthermore, since the
two models show chaotic behaviour as is demonstrated below, it is essential
to understand whether at least the shortest periodic orbits have similar
features; otherwise the corresponding quantum problem is unlikely
to agree with regard to shell structures. In fact, it is the shortest orbits
with smallest period that make the most important
contribution to shell structure in the quantum spectrum\cite{S76}.

We investigate the classical single particle motion for the Hamilton
function (we put the mass equal to unity)
\beq H={1\over 2 }(p_{r }^2 +{p_{\t }^2\over r ^2})+
V_{{\rm WS}}(r ,\t ) \eeq  a
where we use the deformed Woods--Saxon potential $V_{{\rm WS}}(r ,\t )=
V_0/(1+\exp ((r -R(\t ))/d(\t )))$ with $R=R_0(1+\alpha
P_2(\cos \t ))$
and $d=d_0(1+(\nabla R)^2/(2R^2))$ \cite{BM75} where $P_2$ is the second
order Legendre polynomial and $\alpha $ a deformation parameter.
As a second case we consider the classical analogue of the quantum mechanical
Nilsson Hamiltonian(neglecting the spin--orbit term)
\beq H={1\over 2 }(p_{\r }^2 +p_z^2)+
{1\over 2}(\o _{\perp}^2\r ^2+\o _z^2z^2)+
\l (\r p_z-z p_{\r })^2. \eeq b
In both cases there is cylindrical
symmetry. We restrict ourselves to zero value for the z-component of the
angular momentum, that is $p_{\phi }=\dot \phi r^2\sin ^2\t =0$; we have left
out the $\phi $-dependence altogether. We have chosen cylindrical coordinates
$\r $ and $z$ in the second case while the choice $r=\sqrt{\r ^2+z^2}$ and
$\t $ with $\tan \t=\r/z $ is more appropriate in the first case.

We first consider qualitatively the case of spherical
symmetry which means $\alpha =0$ in the first and
$\o _{\perp}=\o _z$ in the second case. Both cases reduce to one
degree of freedom since now $p_{\t }=\dot \t r^2=z p_{\r }-\r p_z$
is conserved. Closed orbits occur if the radial and angular frequencies are
commensurate. Rewriting Eq.(2), for $\o _{\perp}=\o _z$, as
$H=1/2(p_r ^2 +p_{\t }^2/r ^2)+1/2\o _z^2 r ^2+\l p_{\t }^2$ we obtain
closed orbits if $\l p_{\t }/\o _z+1/2=n/m$ with integers $n,m$.
For instance, when $n/m=1/3,1/4,\ldots $ the trajectory forms essentially
a triangle, a square {\it etc.} in the $\r-z-$plane; for $n/m=2/5$ we get
the five star, and so forth. The precise shapes of these geometrical figures
depend on the magnitude of $\l $ in that small values of $\l $ produce
polygons with rounded corners while larger values yield loops at the corners
as is illustrated in Fig.(1). The appropriate scaling of $\l $ is given
by the kinematical constraint between the
energy and the angular momentum which reads $E-\l p_{\t }^2\ge \o |p_{\t}|$.
Here we emphasise that for negative values of $\l $, as used in actual
applications, the value of the angular
momentum $p_{\t }$ is not limited in its absolute value for given energy.
In particular, the relation implies that (for $\l <0$)
if $4 |\l | >\o ^2/E$ there is no restriction at all on $p_{\t }$.
For the Hamilton function of Eq.(1) an appropriate choice of $p_{\t }$ can
likewise lead to polygon orbits such as a triangle, square, pentagon, but
also a five star, and so forth. The corners of the polygons are increasingly
sharp the smaller the value of the diffuseness $d_0$ (or the larger
the value of $R_0$). The angular momentum
$p_{\t }$, and therefore the number of corners of the polygons,
is now limited by the kinematical constraint
$p_{\t }^2\le \max_r [2r^2(E-V_{{\rm WS}}(r ,0))]$.
The plain circle (polygon of infinitely many corners) is possible only for
zero diffuseness where the maximum is reached at $r=R_0$.

Thus a common feature of the two Hamilton functions is, for the trivial
spherical case, the existence of closed orbits of simple geometrical shapes.
Such closed orbits have been observed by other authors for the Woods--Saxon
potential \cite{Ni90,Ma93,Le93} but not, to the best of our knowledge, for
the Nilsson Hamiltonian under consideration. As indicated above, there is
however a crucial difference between the two: the phase space is non-compact
for the second case if $\l <0$. When deformation is invoked both problems
become nonintegrable as $p_{\t }$ is no longer conserved. The symmetrical
periodic orbits discussed above are destroyed. The onset of chaotic motion
can be discerned; for small deformation this happens only in parts of phase
space.

For the Woods--Saxon potential we have chosen only a quadrupole deformation
of the boundary $R$. The deformation of the diffuseness is a consequence of
volume conservation \cite{BM75}. Note that a pure quadrupole deformation of a
cavity ($d_0=0$) yields an integrable case \cite{Ar87} for the bound state
problem. In our case, when $\alpha $ is turned on, the effective deformation
of $V_{{\rm WS}}(r,\t )$ implies higher multipoles of order $2^{2l}, \,
l=1,2,3,\ldots $ since $V_{{\rm WS}}(r ,\t )$ can be expanded in terms of
even order Legendre polynomials. The presence of higher multipoles
is expected to lead to chaotic behaviour \cite{Bl93}. Results of our
analysis confirm this expectation. We have solved numerically
the canonical equations of motion and obtained surfaces of section in the
plane $\r =0$, i.\s e.\s the phase space diagrams displayed in Fig.(2) are in
the $z-p_z-$plane. Our results are general, for demonstration we have
chosen the parameters $V_0=-6$ eV, $d_0=0.74$ \AA ,
$R_0=15$ \AA \ and --3 eV for the energy; this corresponds to the Fermi
level of a cluster with 300 particles \cite{Ni90}. Fig.(2a) represents
sections of three orbits, the initial values of the one in the centre
have been chosen to display the separatrix that separates a periodic
orbit whose trajectory is
displayed in Fig.(3b), and a vibrational mode in the centre of the
$z-p_z-$plane of Fig.(2a). For this particular value of $\alpha $
the motion is still regular in most parts of the accessible
phase space. Another periodic orbit with small stability islands exists
at $z=\pm 17.2$ and $p_z=0$, it is indicated by a pronounced solid dot
and its trajectory
is displayed in Fig.(3a). Only this and the orbit associated with the outer
part of the separatrix can still be traced when $\alpha $ is increased
from 0.1 to 0.16, but only the islands surrounding the latter are still
significant while the ones at the fringes of phase space ($z=\pm 18.7$) have
virtually shrunk to zero; this is illustrated in Fig.(2b).
The onset of chaos now occupies larger parts of the phase space. The fast
decay of the stability islands for only a small range of $\alpha $-values is
remarkable. The centre range in Fig.(2b) is still associated
with quasiperiodic motion. There are many long time periodic orbits which
emerge from the centre. The shortest ones are indicated by crosses and
their trajectories are illustrated in Figs.(3c) and (3d).
The most stable and shortest periodic orbit in Fig.(2b) is associated with
the islands drawn, the trajectory remains that of Fig.(3b).
Hard chaos takes over all of phase space
when $\alpha $ is further increased. The range of $\alpha $-values considered
here is in line with values used by other authors \cite{Fr93}.
We mention that omission of the $\t $-dependence of the diffuseness,
i.\s e.\s putting $d\equiv d_0$, does not change the qualitative picture of
the surfaces of section; it does, however, drastically change an individual
chaotic orbit. All orbits displayed in Figs.(3) emerge from the centre of the
surfaces of section with increasing $\alpha $. While they move outwards in the
section with increasing $\alpha $ their corresponding stability region
shrinks until it disappears.

For the Nilsson Hamilton function of Eq.(2) deformation is invoked by choosing
$\o _{\perp }>\o _z$. The inequality is associated with prolate deformation.
In our context this is the only case of importance since oblate deformation
is related to prolate by interchanging $\r $ and $z$. When $\l =0$ and
$\o _{\perp }>\o _z$ the orbits are Lissajou figures, closed orbits are
obtained for commensurate $\o _{\perp }$ and $\o _z$. Our interest
is directed towards short periodic orbits for $\l \ne 0$ where we actually
focus on $\l <0$ because of its physical relevance. We have investigated the
range of the parameter $b$, defined by $\o _{\perp}=b\o _z$, in the
interval $1.5\le b \le 2$. Fig.(4a) displays the phase space structure
at $\r =0$ where the surfaces of section are taken. Note that, for $\l <0$,
the whole region to the left and right, i.\s e.\s outside the lines
$z =\pm \sqrt{1/(2|\l |)}$, is accessible
as long as the two lines do not intersect with the ellipse which forms the
other part of accessible phase space; if $|\l |$ is sufficiently large to
allow intersection of the lines with
the ellipse, the phase space becomes connected; then the part of
the ellipse which is to the left and right of the lines is inaccessible.
In Fig.(4b) surfaces of sections for $\l =-0.01$ are given for four
different orbits. With the choice of energy $E=50$ and the frequency
$\o _z=\pi /2$ the two lines are outside the ellipse. Recall that
the same pattern is obtained if $E$ and $\l $ are rescaled such that
$E\cdot \l =$constant. With regard to periodic
orbits there is a remarkable similarity between Figs.(2b) and
(4b). In fact, the short periodic orbits are of the same geometrical
shape and occur in similar regions of the surfaces of section.
Compare in particular the rather stable orbit of Fig.(4b) with the
corresponding orbit in Fig.(2b), both have the same trajectory (Fig.(3b))
and their stability  islands are situated in a chaotic region; further the
orbits of Figs.(3c) and (3d) in the centre of the phase diagram which forms
the stable region of Figs.(2b) and (4b); and the orbit
of Fig.(3a) which is situated on the fringes of phase space within a tiny
region of stability. The diagram of Fig.(4b) refers to $b=2$, but the
pattern as described prevails for $1.5<b\le 2$.
For larger values of $|\l |$ hard chaos takes over quickly within the
ellipse, in particular when the two lines enter the ellipse. However, regular
motion prevails outside the ellipse which is the whole area to the left of the
left line and to the right of the right line. These orbits may attain large
values of $\r $ and $z$; also the variation of the angular momentum $p_{\t }$
is unlimited in principle. In this context we stress that the variation
of $p_{\t }$ ranges typically between $-50$ and $+50$ for generic orbits
inside the ellipse for our choice of parameters. This is significant when
considering a corresponding quantum mechanical calculation.

To summarise: the deformed Woods--Saxon potential and the modified Nilsson
model are both used in quantum mechanical models, more recently in applications
to metallic clusters. Their classical analogues do not appear to have much
in common at first sight, yet we have established important similarities
between the two. While this provides confidence in the quantum mechanical
approach where the two models are used for similar purpose, our results
also call for a certain caution. Both models are potentially chaotic with the
degree of chaos depending on the deformation parameter in the Woods--Saxon
potential or on the parameter $\l $ in front of the $\vec l\,^2$-term present
in the Nilsson model. Within the
context of nuclear physics this behaviour is probably of little significance,
since only the lower end of the spectrum is of interest. For metallic
clusters, however, with the much larger particle numbers, a higher range
of the spectrum becomes relevant; in this region chaotic behaviour may well
interfere with the search for shell structure. Our results suggest that
electronic shell structure is not expected to play a major role for the
stability and formation of metallic clusters if a substantial deformation
prevails; however, in previous results, where odd order multipoles were
considered \cite{H94}, the very existence of shell structure was pointed
out for prolate deformation.
Further investigations have to provide clarity on this aspect. Finally we
point out that, in view of the large fluctuations in time of the classical
angular momentum $p_{\t }$, a too severe truncation in the corresponding
(stationary) quantum mechanical calculation could easily invalidate results -
in particular it could conceal the onset of chaos in the quantum spectrum.
\vskip 1cm \noindent
RGN gratefully acknowledges financial support from DGICYT of Spain.

\newpage
\centerline{{\bf Figure captions}}
\vspace{0.5 cm}
{\bf Fig.1} Typical simple periodic orbits for the spherical Nilsson model.
The orbit with the loops at the corners is for $|\l | \simgeq \o ^2/(4E)$
while the other one is for $|\l | = \o ^2/(25E)$.

\vspace{0.5 cm}
{\bf Fig.2} Surfaces of section for the deformed Woods--Saxon potential for
$\alpha =0.1$ (top) and $\alpha =0.16$ (bottom).

\vspace{0.5 cm}
{\bf Fig.3} Typical short periodic orbits for the deformed Woods--Saxon
potential in the $\r -z-$plane. The same shapes occur in the modified
Nilsson model. Figs.(c) and (d) are self-tracing orbits.

\vspace{0.5 cm}
{\bf Fig.4} Surfaces of section for the modified Nilsson model. In (a) two
situations of accessible phase space (shaded) are displayed; left:
$|\l |<\o _z^2/(4E)$, right: $|\l |>\o _z^2/(4E)$.
Four orbits for a value of $\l $ which corresponds to the left of (a) are
displayed in (b).

\end{document}